\documentclass[ltwocolumn]{aastex631}

\newcommand\ec{$\eta$~Car}
\newcommand\hst{{\it HST}}
\newcommand\iue{{\it IUE}}

\newcommand\kms{km~s$^{-1}$}

\newcommand\cmfgen{{\it CMFGEN}}

\newcommand{\Msun}{\hbox{$M_\odot$}}
\newcommand{\Mdot}{\hbox{$\dot M$}}
\newcommand{\Msunyr}{\hbox{$M_\odot\,$yr$^{-1}$}}
\received{tbd}
\revised{tbd}
\accepted{tbd}

\submitjournal{ApJ}

\shorttitle{Eta Carina's Ladder}
\shortauthors{Gull et al.}
\graphicspath{{./}{figures/}}

\begin{document}

\title{Eta Carinae left a curious ladder to climb}

\correspondingauthor{Theodore Gull}
\email{tedgull@gmail.com}

\author[0000-0002-6851-5380]{Theodore R. Gull}
\affiliation{Exoplanets \&\ Stellar Astrophysics Laboratory, NASA/Goddard Space Flight Center, Greenbelt, MD 20771, USA}
\affiliation{Space Telescope Science Institute, 3700 San Martin Drive, Baltimore, MD 21218, US}
\author[0000-0001-9853-2555]{Henrik Hartman}
\affiliation{Materials Science \&\ Applied Mathematics, Malm\"{o}
University, SE-20506 Malm\"{o}, Sweden}
\author[0000-0002-7762-3172]{Michael F. Corcoran}
\affiliation{CRESST \&\ X-ray Astrophysics Laboratory, NASA/Goddard Space Flight Center, Greenbelt, MD 20771, USA}
\affiliation{The Catholic University of America, 620 Michigan Ave., N.E. Washington, DC 20064, USA}
\author[0000-0002-7978-2994]{Augusto Damineli}
\affiliation{Universidade de S$\tilde{a}$o Paulo, IAG, Rua do Mat$\tilde{a}$o 1226, Cidade Universit$\acute{a}$ria S$\tilde{a}$o Paulo-SP, 05508-090, Brasil}
\author[0000-0001-7697-2955]{Thomas Madura}\affiliation{Department of Physics \&\ Astronomy, San Jose State University, One Washington Square, San Jose, CA 95192, USA}
\author[0000-0002-4333-9755]{Anthony F. J. Moffat}
\affiliation{D$\acute{e}$pt. de physique, Univ. de Montr$\acute{e}$al, C.P. 6128, Succ. C-V, Montr$\acute{e}$al, QC H3C 3J7, Canada} 
\affiliation{Centre de Recherche en Astrophysique du Qu$\acute{e}$bec, Canada}
\author[0000-0002-2806-9339]{Noel D. Richardson}
\affiliation{Department of Physics \&\ Astronomy, Embry-Riddle Aeronautical University, 3700 Willow Creek Rd, Prescott, AZ 86301, USA}
\author[0000-0001-9754-2233]{Gerd Weigelt}
\affiliation{Max Planck Institute for Radio Astronomy, Auf dem H\"ugel 69, 53121 Bonn, Germany}

\begin{abstract}
Eta Carinae underwent the Great Eruption in the 1840s and a Lesser Eruption in the 1890s. Its apparent spectrum, modified by intervening ejecta, the Homunculus and Little Homunculus, continues to evolve but contains information pertaining to events in the 19th century.  The LOS spectrum contains narrow absorption velocities, from $-$122 to $-$1665 \kms: rungs of a broken ladder caused by shells formed by the interacting winds. Estimated shell origin dates correlate with origin dates of expanding emission structures preceding the Great Eruption. The LOS absorption velocities extend the record post Great Eruption to the Lesser Eruption.  We suggest that these shells originated from a binary merger within a triple system. Shells formed not only from periastron passages of the current secondary, but also from ear-like extensions preceding and following the periastron event. Additional models need to be considered.

\end{abstract}


\section{Introduction \label{sec:intro}}

Eta Carinae (\ec) and  its ejecta have intrigued astronomers for two centuries. Observers from several southern stations recorded the apparently erratic behavior of $\eta$ Argus, as it was known in the early 1800s, but paid closer attention as it brightened in the 1840s to rival Sirius, then faded to visual obscurity \citep{Frew04, SmithFrew11}. This event is known as the Great Eruption. A much fainter but significant brightening, known as the Lesser Eruption, was recorded in the 1890s. 

Ejecta from these two eruptions formed the Homunculus, a dusty, bipolar shell expanding at $\approx$650 \kms\ in the polar regions \citep{Smith03}  and the Little Homunculus, an ionized, bipolar shell nested within the Homunculus, expanding at $\approx$170 \kms\ \citep{Ishibashi03}. The total ejected mass, most of which appears to reside in the skirt region between the two lobes,  is estimated to exceed 45 \Msun\ \citep{Morris17} although other estimates are less \citep{Smith18a, Davidson97}.

Today \ec\ is a highly eccentric binary whose period, 5.54 years, is  based upon a nebular high-ionization state ([\ion{Ne}{3}], [\ion{Fe}{3}], [\ion{Ar}{3}], \ion{He}{1}, and \ion{He}{2}) modulated by a short,  low-ionization state ([\ion{Fe}{2}]) \citep{Damineli96, Damineli97, Teodoro16}. The repeating X-ray light curve confirmed the binary orbit \citep{Corcoran01a}. \cite{Pittard02} first estimated the wind properties of both the primary and secondary. Refinement of the wind parameters with further observations has led to the following:
M$_A \approx$ 100 \Msun\ \citep{Strawn23}, \Mdot$_A =$ 8.5$\times$10$^{-4}$, \Msunyr \citep{Madura10}, V$_{tA} =$ 420 \kms\ \citep{Groh12} , M$_B >$ 60 \Msun\ \citep{Strawn23}, \Mdot$_B =$ 10$^{-5}$\Msunyr, V$_{tB} =$ 3000 \kms\  \citep{Pittard02}.

The visible brightness of \ec\ has increased since the 1940s \citep{Damineli21}, but the Homunculus, used as an infrared calorimeter to measure dust re-emission of absorbed starlight, suggests no change in the binary luminosity over the past half century \citep{Mehner19}. 

\cite{Hillier01, Hillier06} noted that the measured H$\alpha$ equivalent width in the LOS was twice that measured of the scattered starlight off the Homunculus. The directly measured H$\alpha$ equivalent width was much weaker than that expected from \cmfgen\ modeling. They suggested that the continuum-emitting core and a portion of the H$\alpha$-emitting, extended wind of \ec-A must have been occulted in our LOS.  

A logical concept would be that a foreground dusty structure with a well-defined edge  partially obscured \ec. The studies of \cite{Damineli21} and citations therein show that the H$\alpha$ equivalent width in the LOS has decreased over the past two decades,  approaching the stable equivalent width measured in foreground Homunculus-scattered starlight. The occulter has been dissipating over the past few decades.

The half-century infrared stability and the  X-ray light curve stability over the past five cycles \citep{Espinoza22} led to our understanding that the binary properties have remained stable in recent decades but the ejecta in our LOS have been evolving \citep{Gull23}.

The causes of the two eruptions have been a mystery. The extreme abundance of nitrogen relative to carbon and oxygen and the classification of \ec-A as a luminous blue variable (LBV) suggested to some that the companion star is an evolved WN star that might have undergone a near-supernova event, but this is unlikely as \ec-B survived. Certainly the lower limit, $>$ 60 \Msun, provides the case for overproduction of nitrogen at the expense of carbon and oxygen \citep{Ekstrom12}and implies a less-evolved H-rich WNh star (like the three nearby luminous WNh stars WR22, WR24 and WR25, the first and third of which are massive binaries - \cite{LenoirCraig22,Gamen06}) instead of an evolved H-poor classical WN star. 

The terminal velocity derived from the X-ray spectrum is 3000 \kms\ \citep{Pittard02}. The energetics of the Great Eruption are comparable to a supernova event but the energetics of the Lesser Eruption is considerably less \citep{Davidson97}. However no definitive observation has demonstrated that the secondary is a WN star, except possibly for the presence of a diluted \ion{He}{2}~$\lambda$4686 emission line outside periastron that moves in antiphase with the primary LBV \citep{Strawn23}. \ion{He}{2} $\lambda$4686 is the brightest emission-line in WN stars and some extreme, hot Of stars. Currently no observation has resolved the two stars and determined their properties independently. 

An alternative suggestion is that the two events were the result of a merger within a triple system that evolved to a double system. Two  models of such a system applied to \ec\ were provided by \cite{Portegies16} and \cite{Hirai21}. We focus on the model by \cite{Hirai21} and discuss how a series of narrow velocity components, noticed by \cite{Gull06}, might provide crucial information about the possible merger. In an analogy of a ladder, these velocity components appear to be rungs that record changing information about the merger beginning before the Great Eruption, extending across a chaotic period, and settling across the Lesser Eruption as the system evolved from a triplet to the present day binary.

Recently, \cite{Morse24} measured \hst\ imagery and spectroscopy of the expanding Homunculus in the light of \ion{Mg}{2}~$\lambda\lambda$2796, 2803 and found systematic expansion velocities corresponding to ejection dates of 1847 extending as early as 1760. We will demonstrate that their observations are compatible with our LOS absorption velocities measured by \cite{Gull06}.

Our effort is to gain insight on the events extending across the nineteenth century. We have three sets of measurements that we will attempt to associate: the intermittent visual magnitude estimates by observers during the event, the expansion of the Homunculus measured from two epochs of direct images with the {\it Hubble Space Telescope (HST)}\footnote{Based on archived observations made with the NASA/ESA Hubble Space Telescope, obtained at the Space Telescope Science Institute, which is operated by the Association of Universities for Research in Astronomy, Inc., under NASA contract NAS5-26555.}, and a series of high-dispersion, NUV spectra of \ec\ containing a plethora of narrow absorption lines in our LOS.

The visual magnitudes provide a brightness measure of the stellar and nebular system across the nineteenth century. The direct images enabled an averaged expansion in the sky plane projecting back to the Great Eruption. The NUV \hst/STIS spectra reveal absorption systems in the LOS pencil beam  which extend out of the sky plane sampling the three-dimensional (3D) Homunculus tilted at $\approx$ 45$^o$. 

We suggest that this ladder of absorption velocities was created in the 19th century and provides important clues to the physics that led to the two great eruptions. Unfortunately astronomical spectroscopy did not exist at the time of the Great Eruption. Only visual photometry was possible. However we have this record of shell absorption velocities that may provide insight on what happened with \ec\ and possibly an explanation of the observed pseudosupernovae seen in distant galaxies. 

The observational data recording the three dozen velocity components \citep{Gull06} are summarized in section \ref{sec:obs}. Linking the velocity components to the historical events of the nineteenth century and the model of \cite{Hirai21} is in section \ref{sec:dis}. Conclusions and some suggested future observations are in section \ref{sec:con}.

\section{The observations}\label{sec:obs}

\begin{figure}[ht]
\includegraphics[width=8.5 cm]{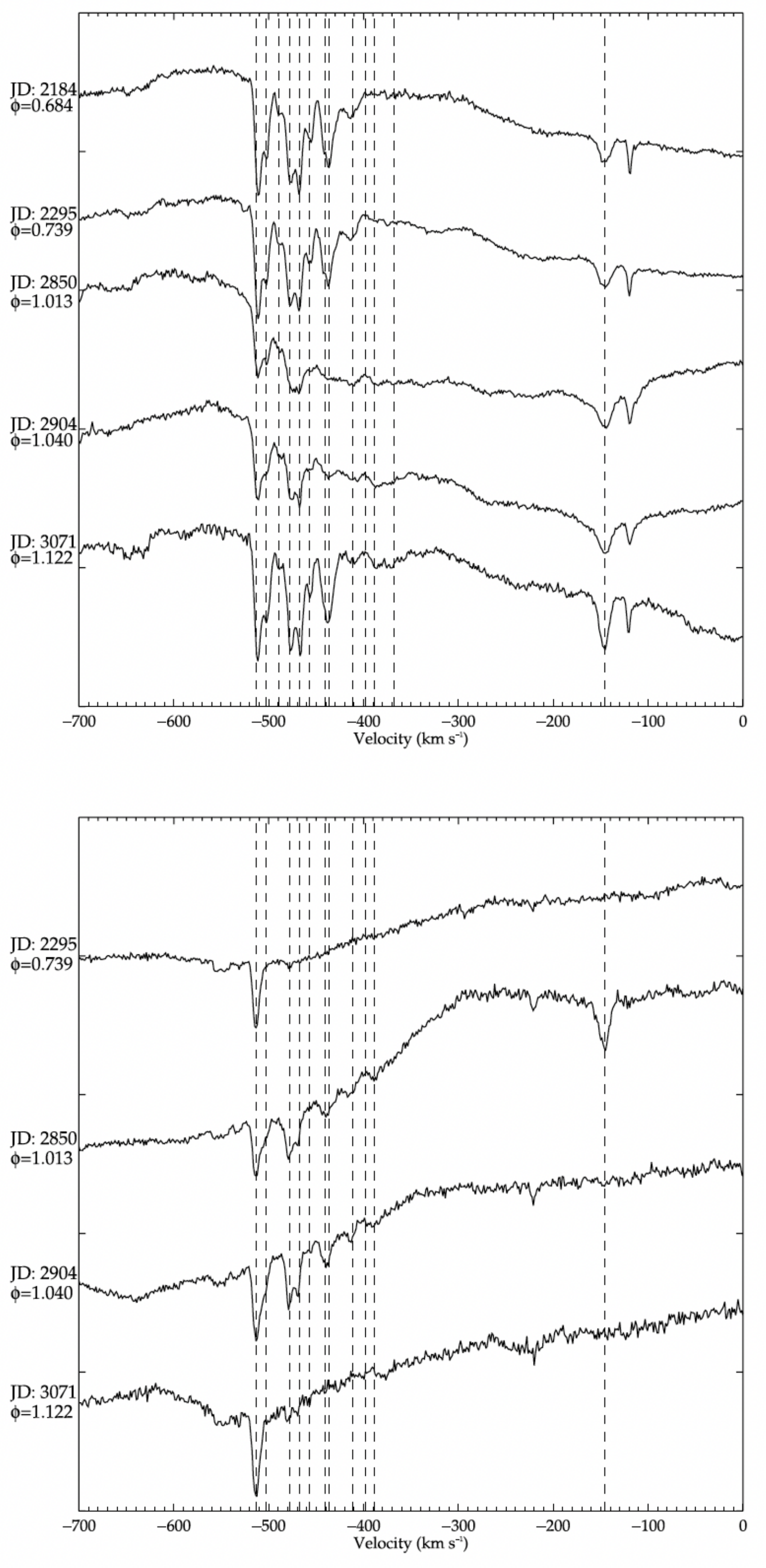}
          \caption{{\bf Examples of intermediate velocity components that varied with spectroscopic phase. Top:} \ion{Mn}{2} $\lambda$2934    weakened during the low-ionization phase because fewer NUV photons led to excitation of upper energy levels. {\bf Bottom:} \ion{Ti}{2} $\lambda$3080 strengthened during the minimum as the Lyman continuum, which ionizes Ti$^+$ to Ti$^{++}$, originating from \ec-B, dropped across the periastron passage (reproduced from \cite{Gull06}).
\label{fig:spect}}
\end{figure}

Numerous UV spectra had been previously recorded of \ec\ with the International Ultraviolet Explorer ({\iue}) in campaigns aimed at identifying the complex spectral features and potentially the source of variability. \cite{Viotti89} provided an impressive line identification list, given that the \iue\ spectral resolving power was R $= $ 10,000 with an oval aperture, 10\arcsec\ $\times$
 20\arcsec. The \iue\ aperture, dependent upon aperture position angle, included varying contributions of nebular-scattered light from the bipolar 10\arcsec\ $\times$ 18\arcsec\ Homunculus with expansion velocities in the range of  $+/-$650 \kms\ \citep{Smith06}. Hence the recorded spectra of \ec\ included velocity-smearing components leading to considerable confusion in identifying complex absorptions. Until the mid-2010 era, the nebular-scattered component dominated the 3\arcsec\ diameter core, including \ec, found by \cite{Damineli21}. 

 The ten-fold increase in spectral resolving power provided by the \hst/STIS high-resolution NUV\footnote{We define the near ultraviolet (NUV) to extend from 2400 to 3100\AA, the MUV as 1700 to 2400\AA\ and the far ultraviolet (FUV) from 1200 to 1700\AA, consistent with the spectral intervals covered by the STIS. The EUV extends shortward of 1200\AA.} echelle mode, the much greater dynamic range of the NUV Multi-Anode Multi-Array detector (MAMA), and the greatly improved NUV angular resolution, 0\farcs07, of \hst, compared to the 10$\arcsec\times\ 20\arcsec$ spatial sampling of \iue, led to detection of thousands of low-ionization, metal absorption lines with multiple velocity components in our LOS \citep{Gull06}. Indeed the bulk of the UV scattered light originates from the 1\arcsec\ diameter region centered on \ec. The velocity dispersion across that region easily smears out the nebular absorption components. Hence the diffraction-limited spatial resolution of \hst\ in the NUV is critical to the study of the LOS absorptions.
 
NUV \hst/STIS  echelle spectra, recorded between October 2001 and March 2004, of \ec\ provided an unexpected super abundance of very narrow absorption lines on top of an already complex spectrum of broad P~Cygni wind line profiles \citep{Gull06}. Hundreds of absorption line complexes were resolved by the \hst/STIS capabilities of high resolving power, R $= \lambda/\delta\lambda =$ 110,000, combined with the high angular resolution, $\delta\theta =$ 0\farcs07 ($ =$ 170 au at the adopted distance of 2300 parsecs; the projected binary orbit is thought to be about 30 au, $\approx$ 12 milli-arcseconds). 

Far  ultraviolet (FUV)  spectra recorded in the same time interval with the \hst/STIS using a lower resolving power, R~$=$ 46,000, revealed over 800 absorption lines of H$_2$ \citep{Nielsen05a} all at $-$513 \kms. Additional metal absorption lines were identified in the FUV most notably at $-$513 and $-$146 \kms. The decreased resolving power, the H$_2$ absorptions, and the broad wind lines complicated clear separation of the narrow absorption lines, but still the FUV spectra provided crosschecks on the information in many cases as  especially in the more recent FUV observations recorded with R=100,000 \citep{Gull21a,Gull22,Gull23}.

While all of the H$_2$ absorptions originated from the Homunculus at $-$513 \kms, the metal absorptions originated from multiple  velocity components \citep{Gull05, Gull06}. The strongest narrow absorption lines at $-$513 \kms, were identified with the Homunculus, but many strong, but broader absorption lines came from a velocity system at  $-$146 \kms, associated with the Little Homunculus discovered and mapped by \cite{Ishibashi01a}.

\begin{table*}
\tablenum{1}
\centering
\caption{Circumstellar Absorptions ( from \cite{Gull06})} \label{tab:lines}
\begin{tabular}{lrll}
\hline
Velocity&$\delta v^a$&Neutrals and Ions Observed within Velocity Component$^b$&Inferred Date\\  
\kms&\kms& &of origin\\
\hline
$-$122$^c$  &--&   Mg I, Mg II, Cr II, Mn II, Fe II, Ni II&1897.9\\
$-$146$^d$ &24&  He I, Mg I, Mg II, Ti II, V II, Cr II, Mn II, Fe II, Co II, Ni II&1894.8\\ 
$-$168$^c$ & 22& Mg I, Mg II, Cr II, Mn II, Fe II, Ni II&1891.9\\
$-$185$^c$  &17&   Mg I, Mg II, Cr II, Mn II, Fe II&1889.7$^e$\\
$-$323$^f$  &138&  Mg II, Mn II, Fe II&1871.6\\
$-$335$^f$ & 12&Mg II, Mn II, Fe II&1870.0\\
$-$345$^f$ &10&  Mg II, Mn II, Fe II&1868.7\\
$-$359$^f$& 14&Mg II, Mn II, Fe II&1866.9\\
$-$369&10&   Mg II, Mn II, Fe II&1865.5\\
$-$389 &20& He I, Mg II Mg II, Ti II, Mn II, Fe II&1862.9\\
$-$398 &9& Mg I, Mg II, Mn II, Fe II&1861.8\\
$-$412 & 14&Mg I Mg II, Ti II, Cr II, Mn II, Fe II&1859.9\\
$-$436& 24&Mg I, Mg II, Ti II, V II, Cr II, Mn II, Fe I, Fe II&1856.8\\
$-$441 &5&Mg I, Mg II, Ti II, V II, Cr II, Mn II, Fe I, Fe II&1856.1\\
$-$457& 16&Mg I, Mg II, Ti II, Cr II, Mn II, Fe II&1854.0\\
$-$468&11& Mg I, Mg II, Ti II, V II, Cr II, Mn II, Fe I, Fe II&1852.6\\
$-$478 &10&Mg I, Mg II, Ti II, V II, Cr II, Mn II, Fe I, Fe II&1851.3\\
$-$489 &11&Mg I, Mg II, Mn II, Fe II&1849.82\\
$-$503& 14& Mg I, Mg II, Ti II, V II, Cr II, Mn II, Fe I, Fe II&1848.0\\
$-$513$^g$ &10& Na I$^h$,   Mg I, Mg II, Si I, Sc II,   Ti II, V II, Cr II, Mn II, Fe I, Fe II, Ni I, H$_2$&1846.7$^i$\\
$-$527&14&  Mg I, Mg II, Mn II, Fe II&1844.8\\	
$-$537 &10&  Mg II, Mn II, Fe II&1843.5\\
$-$553 &16& Mg I, Mg II, Mn II, Fe II&1841.4\\
$-$561 &8& Mg I, Mg II, Mn II, Fe II&1840.4\\
$-$572 &11&   Mg I, Mg II, Mn II, Fe II&1839.0\\
$-$587& 15&Mg I, Mg II, Mn II, Fe II&1837.0\\
$-$604 &17& Mg II&1834.8\\	
$-$696 &92&Mg II&1822.7\\
$-$705&  9&Mg II&1821.5\\
$-$712&7& Mg II&1819.4\\
$-$732 & 20&  Mg II&1818.0\\
$-$833&101 &   Mg II&1804.7\\
$-$1038$^j$  &205&  Mg II&1777.9\\
$-$1050$^j$&12&  Mg II&1776.3\\
$-$1064$^j$  &14& Mg II&1774.5\\
$-$1074$^j$&10 & Mg II&1773.2\\
$-$1640$^k$&566&C IV, Si IV, Al III& 1699.0\\
$-$1665$^k$&15&C IV, Si IV, Al III & 1695.7\\
\hline
\end{tabular}\\
$^a$ Increment in velocity relative to the previous entry. 
$^b$ Two or more lines led to velocity component identification \citep{Gull06}. Typical absorption widths were 2.2 \kms\ for $-$513 \kms\ and 5.5 \kms\ for $-$146 \kms.  Absorption widths for all velocity components more negative than $-$300 \kms\ were 2.2 \kms. 
$^c$Identified during the 2003.5 minimum spectra only.
$^d$  Unique  Little  Homunculus  velocity  component  that  is  well  isolated  in velocity.
$^e$  Based upon photometric increase and plateau \citep{SmithFrew11}.
$^f$ Identified in 2004 March 4 postrecovery spectrum only.
$^g$ Unique Homunculus velocity component that is well isolated and has many absorption lines, including H$_2$ in FUV spectra \citep{Nielsen05a}.
$^h$ Single line identification, confirmed by transitions at VLT UVES optical wavelengths.
$^i$ Based upon Homunculus expansion  \citep{Smith17}. 
$^j$ Confirmed in Si II, Si IV, Al II, and Al III in the MUV spectrum \cite{Nielsen05a}.
$^k$ Two additional absorption velocities identified by \cite{Nielsen05a}
\end{table*}
The bulk of the absorption lines came from additional, weaker velocity systems. A total of 38 velocity components were identified ranging from $-$122 to $-$1665 \kms\ (Table \ref{tab:lines}). The measured velocities are spaced at intervals varying from 5 to 24 \kms\ but with several significant, larger gaps. 

Numerous examples of multi-velocity, absorption profiles are presented in \cite{Nielsen05a,Gull06}. Strengths of the velocity components were found to vary within the 5.54-year binary orbital phase primarily affected by changes in the FUV and EUV fluxes as the interacting winds shift from the high- to low-ionization states. 

Strong absorptions of \ion{Mn}{2}~$\lambda$2934 weakened during the low-ionization phase as NUV photons leading to excitation of upper energy levels decreased (Figure \ref{fig:spect}, Top). 

Strong absorptions of the  \ion{Ti}{2} $\lambda$3080 at or near $-$146 \kms\  briefly appeared during the months-long low ionization state across periastron, but were absent  across the high ionization state which extends across five years of the binary period (Figure \ref{fig:spect}, Bottom). Ti$^+$, with its ionization potential, 13.58 eV, just below the ionization potential of hydrogen, indicates the presence or absence of UV photons shortward of the Lyman continuum. Strong variations in the four velocity systems between $-$122 and $-$184 \kms\ occurred between the high- and low-ionization states. Similar variations were recorded in more recent, ground-based studies of Na~D absorption over the last three orbital cycles \citep{Pickett22}. 

Absorption velocities identified between $-$603 and $-$527 \kms\ required considerable effort as the very narrow features are superimposed upon the rising edge of a broad absorption. Examples are absorptions arising from \ion{Fe}{2} transitions with lower energy levels about 1 eV above the ground state. Patterns of very sharp (FWHM $\approx$ 2 to 5 \kms) absorptions with velocities more negative than $-$513 \kms\ were repeatedly identified by \cite{Gull06} in the multiple echelle spectra displayed in their Figure 8. Indeed the strengths of these transitions prove to be so strong that broad, totally saturated absorptions exist between $-$513 and $-$146 \kms. 

No single absorption line provides measurable profiles for all velocities. Less abundant metals such as titanium provide transitions with moderate f-values leading to unsaturated velocity profiles for systems with velocities less negative than $-$513\kms. The much more abundant iron in singly-ionized state has transitions with strong f-values around 2500\AA\ that lead to totally saturated profiles ranging from $-$513 to $-$380 \kms\ and $-$210 to $-$100 \kms (see the many examples in Figure set 9 in \cite{Gull06})

Two additional high-velocity absorption components, $-$1640 and $-$1665 \kms\ were noted in the MUV by \cite{Nielsen05a}. The rest of the absorption components are at $-$1074 or closer to the velocity of \ec\ and occurring at much more frequent intervals.  Likely these two absorption velocities are associated with the high velocity emission structures noted by \cite{Kiminki16} and the ghost nebula originally identified by \cite{Currie02}.

\section{Discussion}\label{sec:dis}

\cite{Gull05} analyzed absorption profiles of singly-ionized metals in the two strongest velocity components, $-$513 and $-$146 \kms. They found the lower levels from which the absorptions originated, were thermally populated. Across the high-ionization state, the $-$146 \kms\ system metals had a kinetic temperature of 6400~K while the kinetic temperature of the $-$513 \kms\ system was 760~K. These temperatures infer that, as expected, the higher-velocity  structures are thermally cooler at greater distances. 

Many atomic and ionic species, identified in the NUV spectra as noted in Table \ref{tab:lines}, were identified in the low-velocity components. The number of species decreased with more negative velocities with the noticeable exception of $-$513 \kms\  and nearby velocities.  At most velocities more negative than $-$604 \kms, only absorptions in \ion{Mg}{2} were identified. Six velocity components at or more negative than $-$1038 \kms are exceptions as they are seen  in  multiply-ionized states despite being more distant.  

Considerable changes occurred between the high-ionization and the low-ionization states \citep{Gull06}. The kinetic temperature of the $-$146 \kms\ metals dropped from 6400~K to 5000~K from the high- to low-ionization state in late 2003, then increased by March 2004. While change in the kinetic temperature of the $-$513 \kms\ shell was not noticeable, virtually all of the H$_2$ absorptions disappeared across the low-ionization state due to the significant drop in the STIS FUV flux \citep{Gull22}. Little or no radiation shortward of Ly$\alpha$\ was present. Hence the H$_2$ relaxed to ground state and H$_2$ destruction temporarily decreased.

These velocity systems do not originate from individual clumps in our LOS, but from spatially-extended, clumpy shells as demonstrated by \cite{Gull23}. Long slit spectra with the \hst/STIS in the NUV, with the G230MB grating and CCD detector (R=10,000), were recorded during the high-ionization states in cycles 10 and 13. While not at the resolving power of 100,000, the spatially resolved (0\farcs1 limited by the CCD format) spectra demonstrated that 1) the merged velocity systems extended across the nebula, centered on the position of \ec, and 2) many velocity components disappeared due to the ten-fold increase of FUV flux in the high states between cycles 10 and 13. 

The velocity systems, while clumpy in character as noted by the changes in equivalent widths across the high-ionization state of cycle 10, form shells extending across significant portions of the foreground Homunculus. Hereon we will address these absorption systems as shells formed across multiple periastron passages. 

This series of multiple absorption velocities, effectively form a ladder with rungs defined by individual absorption velocities. This velocity ladder records  wind velocities that decreased across a  very extended time interval, namely across the nineteenth century and  offers clues to understanding the phenomena that caused the two major outbursts. 

In the following subsections, we will now build the case that associates the absorption velocities with shells. We initially sketch how the shells formed (subsection \ref{sec:shells}) and suggest a mechanism that led to the narrowing of individual velocity profiles (subsection \ref{sec:form}). In subsection \ref{sec:change} we describe how the shells have evolved over the past two decades, which provides insight on how the shells formed and survived throughout the nineteenth and twentieth centuries.  We provide connection to the Great and Lesser Eruptions, which then leads to possible dates when the individual shells formed (subsection \ref{sec:date}).  In subsection \ref{sec:period} we try to establish an orbital period between the two eruptions. Subsection \ref{sec:few} mentions the possibility that the different absorption velocities might have been formed by a few discrete ejections. Subsection \ref{sec:linear} adjusts the velocities to form a linear change in velocity with time. Subsection \ref{sec:dac} considers individual wind flows seen in other OB stars. We finally discuss in subsection \ref{sec:scenario} how the velocity ladder may follow a merger scenario which appears to be the more appealing explanation but  still has shortcomings . 

\subsection{ How did the shells form and how did they survive until very recently? \label{sec:shells}}

Colliding winds of massive binaries can lead to continuously-formed structures in the form of equidistant, expanding shells. These winds are modulated by  the physical conditions and changes thereof. The very noticeable spiral structures surrounding WR140 are the result of colliding winds leading to dust formation under optimal physical conditions  \citep{Lau22}. Nearly eighteen nested spiral shells of dust were imaged expanding at uniform velocity indicating the colliding winds have been stable across  eighteen orbits of the binary WR140. At specific phases of the highly-ellipitical (like \ec\ with e $=$ 0.9) binary orbit, dust formed in abundance.

Similar evidence of  stable colliding winds from the current binary of \ec\ was found by \cite{Teodoro13} in the form of forbidden emission from three nested hemispherical shells located within 0\farcs5 (1150 au) of \ec. From imagery in [\ion{Fe}{2}] and [\ion{Ni}{2}] mapped at intervals from 2009 to 2013, they found the shells were expanding at 470 \kms, slightly more than the terminal velocity of \ec-A, 420 \kms, previously determined by \cite{Groh12}. The much faster, lower density, secondary wind collides with the primary wind which is compressed and accelerated from 420 to 470 \kms.

3D hydrodynamical models suggest that these shells were formed within the primary wind as the secondary with its wind passed through each periastron passage, thus compressing a portion of the primary wind into thin, outwardly-moving shells modulated by the low-density, fast secondary wind \citep{Madura12B, Madura13}. 

Additional shells, expanding at 470 \kms, were not detected at greater distances. Slow-moving, much more-massive ejecta, exemplified by the Weigelt clumps B, C and D \citep{Weigelt86, Weigelt12}  provided a barrier expanding at $-$40 \kms. 

Apparently a dusty occulter, now nearly dissipated \citep{Damineli23,Gull23}, is part of an extensive slow-moving structure that has protected the multiple shells originating in the nineteenth century from before the Great Eruption to shortly after the Lesser Eruption. Multiple structures, seen in forbidden emission, including the Weigelt clumps,  with expansion velocities less than 100 \kms, were mapped by \cite{Gull16} within the central arcsecond surrounding \ec.
 
Much of the ejecta was compressed. The gases  cooled, then formed molecules and  dust. Due to the paucity of carbon and oxygen, unusual dust formed, possibly influenced by nitrogen-dominated chemistry, leaving considerable metals in atomic or ionic state. Detection of \ion{Sr}{2}, \ion{Sc}{2} and \ion{V}{2}, both in absorption and forbidden emission in ejecta close to \ec\ is unique \citep{Hartman04,Hartman05}. No previous records of absorption lines from these ions had been identified in the ISM. Thus the carbon- and oxygen-deprived chemistry of the dust formed by \ec\ appears to be unique.

Over the past century, multiple shells, separated by ever-decreasing velocity intervals, have been expanding outward from \ec. The molecules and dust, within the slow-moving ejecta close to \ec,  gradually are being destroyed leading to increasing amounts of ionizing radiation. The neutral and singly-ionized gas shells have now become doubly-ionized shells with absorptions well into the UV beyond Ly$\alpha$. An extreme example is the nearly complete destruction of the $-$513 \kms\ H$_2$ between 2004 and 2016 in the Homunculus   \citep{Gull22}.

In the long term, the Homunculus will continue to expand, the dust and molecules in the shells  will be destroyed and the multiply-ionized shells will increasingly become invisible to the observer because most resonant transitions of multiply-ionized metals are located below Ly$\alpha$. The high dispersion UV observations of the ejecta over the past two decades record major changes of the ionization state of the Homunculus in our LOS.

The observational records show that \ec\ faded  to seventh visual magnitude in the late 1890s through the late 1930s, then began to gradually brighten \citep{SmithFrew11}. Today, \ec\ is again a naked eye object recently approaching fourth magnitude \citep{Damineli19}. While some suggest that the 1940s brightening was yet another event \citep{ Abraham14}, most are convinced that this was the result of gradual dissipation of the Homunculus originating from the nineteenth century and continues through the present \citep{Smith18c, Damineli23, Gull23}.

\cite{Morris17} found that the peak infrared emission originates from a region about two arcseconds north of \ec. They estimated the total mass inferred from this dust emission to exceed 40 \Msun. Such is consistent with most of the slowly-moving ejecta flowing outward on the far side of the binary orbit in the general direction of the current periastron as seen by the observer. 

\subsection{A mechanism leading to narrowing of the absorptions}\label{sec:form}

The result of colliding winds is a velocity distribution about a momentum-balanced velocity broadened by thermal and turbulent contributions. Yet the absorption components in the ladder are extremely narrow compared to the wind absorption profiles (absorptions extending to $-$500 \kms): shells with the LOS velocities more negative than $-$383 \kms\ are narrower, fwhm at 2.2 \kms, than the $-$183 to $-$136 \kms\ components associated with the Lesser Eruption, fwhm at 5 \kms \citep{Gull06}.

How did the individual absorption profiles narrow in width? Some insight comes from comparing the multiple shells around WR140 and the ladder shells. As the binary orbital plane approaches face-on (inclination $=$ 119$^o$ \citep{Thomas21}), the dust spiral around WR140 appears to be uniformly spaced \citep{Lau22}, indicating that the balance of the colliding winds and  the ionizing radiation has been constant for the last 18 orbits. This in turn suggests that the physical conditions leading to the dust formation in WR140 have not varied significantly over those many orbits, which exceeds a century in time. 

In contrast, the absorption ladder detected in the LOS of \ec\ is a record of a long-term decrease in wind velocity over many cycles. Over that long term, likely \ec-B, given that  WNh stars have a long lifetime ($\approx$ 10$^6$ years), had a constant wind over the past few centuries. 

Today, the wind of \ec-A is very extended. \cite{Teodoro16} mapped three hemispherical shells in [\ion{Fe}{2}] and [\ion{Ni}{2}] that by position and velocity were shown to be formed over the three low-ionization states when \ec-B plunged into the extended primary wind across the periastron passage. For a few months the primary wind flowed into the surrounding volume while \ec-B formed a highly-ionized volume separating a portion of the primary wind forming into a partial shell. When \ec-B emerged from the primary wind, its faster, less -massive wind cleared out a complimentary hemisphere. The compressed, thin shell, highly-ionized by \ec-B flowed outward, analogous to the highly-ionized wind shells of WR140. However, as noted above, only three shells survive before being absorbed by slower moving clumps beyond  \citep{Teodoro16, Gull23}.

Narrowing of wind absorption lines, generated by gas flowing from a bright spot, has been demonstrated repeatedly for OB stars as done by \cite{Kaper97} in monitoring programs with the \iue. A bright spot, originating as a perturbation, causes a faster flow than the rest of the wind initially, that turns into a slower flow as the star rotates and the cooler stellar surface under it can no longer support such a fast flow, resulting in a broad absorption in the P Cygni profile. Extreme ultraviolet radiation (EUV) accelerates the highly-ionized, absorbing gas. A few strong FUV resonant or near-resonant absorptions in \ion{Si}{4} or \ion{C}{4} appear at low, broad velocities, then accelerate to higher velocities, narrowing in dispersion and terminating near the maximum (terminal)  velocity possibly limited by velocity crowding. 

Sufficient EUV originates from most OB stars to multiply ionize the escaping wind and by momentum transfer accelerates slower moving ions up to a terminal velocity where the radiation is totally absorbed.

The current colliding winds of \ec\ create two very different structures dependent upon the binary separation. For most of the orbit, the EUV radiation of \ec-B multiply-ionizes and compresses the wind of \ec-A which results in a highly-ionized, low-density set of shells not readily detected. Across periastron, \ec-B dives into the extended atmosphere of \ec-A which results in a highly-compressed, high density shell.

The greatly extended envelope of \ec-A absorbs most of the EUV from \ec-B with the result of a compressed shell  with metals mostly in singly-ionized state. Three shells, seen in [\ion{Fe}{2}] and [\ion{Ni}{2}] were tracked by \cite{Teodoro16} and shown to originate from the three previous periastron passages.

Today, the EUV, FUV originating from the secondary companion, assumed not to have changed very much since the 19th century, ionizes most of the gas from each cycle outward to the dissipating, occulting structure, thought to be located less that 2000 AU from \ec\ \citep{Gull23}. Beyond the occulter, resonant and near-resonant transitions of neutral and singly-ionized metals are much more numerous throughout the MUV and FUV compared to the EUV. The observed MUV spectrum of \ec\ is almost a desert due to saturated metal absorptions by the multiple shells. The NUV, where fewer, weaker resonant transitions reside, repeatedly show patterns of narrow-line absorptions for individual resonant lines, defined by velocities of the individual shells.

 Each absorbed photon leads to incremental momentum transfer. The accumulation of hundreds of resonant and near-resonant transitions in the MUV and NUV narrows the absorption profile at a terminal velocity, usually somewhat larger than the initial terminal velocity of the material leaving the extended envelope. The net result is a narrowed absorption profile for each individual shell. Note that absorptions from the four shells with least negative velocities (closest to \ec) have significantly broader absorption profiles.
 Their large column densities and higher kinetic temperatures have led to absorption profiles that have narrowed less that the older, more distant shells with more negative velocities.

The EUV originating from \ec\ in the nineteenth century  would have been even more strongly absorbed by the greatly extended envelope due to the Great Eruption. The envelope surrounding the merging binary would have puffed up considerable, possibly with decreasing terminal velocity over the merger event. Observers commented that \ec\ had a very red color \citep{Smith06,SmithFrew11}. The light echo spectrum, that has been tracked moving across relatively distant reflection nebulae, appears to be close to that of an F star \citep{Rest12}. The primary absorbers within the shells would be neutral and singly-ionized metals, not ionized hydrogen and helium nor multiply-ionized metals. 

With the EUV depleted, the metals in the shells would be in neutral and singly-ionized states.  Hundreds of resonant and near-resonant lines provide considerable momentum transfer greatly narrowing the velocity dispersions of each shell. Radiation transfer by photons, long ward of Lyman continuum, to the metals would be strongest in the most recent shell, accelerating the gas to increased velocity, but scaled by d$^{-2}$ and most likely limited by a decreasing terminal velocity.

Hence the shell would be accelerated to a maximum velocity very quickly. The width of the velocity absorptions is systematically broader for the lowest four velocity components associated with the Little Homunculus which suggests either that the gas column density was quite large or more likely the FUV/NUV flux was blocked by material moving more slowly, i.e. structures including the Weigelt clumps.

\subsection{Changes across the past two decades} \label{sec:change}
The inventory of these absorption velocities was accomplished with \hst/STIS echelle spectra recorded in the NUV between 2000 and 2004 \citep{Gull06}. Variations in the strength of the absorptions increased notably as the binary moved from the high-ionization state to the low-ionization state  as the FUV (and implied EUV) flux, originating from \ec-B, dropped across the periastron passage while the companion was immersed within the extended envelope of \ec-A for a few months \citep{Gull23}.

Greater changes occurred between 2000 and 2018 as the apparent FUV flux seen by the shells increased nearly ten-fold due to the gradual dissipation of a dusty occulter located close to \ec\  \citep{Damineli23, Gull21a, Gull23}. The increasing FUV radiation (as a proxy to the EUV below Ly$\alpha$) led to increasing ionization starting with the shells closest to \ec, extending increasingly to the more distant shells. Virtually all neutral and singly-ionized absorption lines disappeared in the FUV \citep{Gull21a}. Only moderate, long slit spectra (R=10,000) were obtained in the NUV since 2004. These spatially-resolved spectro-images showed that the line absorptions decreased in strength in our LOS and extended across a portion of the foreground Homunculus lobe \citep{Gull23}.

\cite{Hillier01,Hillier03} and \cite{Damineli23} hypothesized, then demonstrated that the increase in flux extending from the visible to the FUV is caused by a dissipating occulter in our LOS. \cite{Damineli23} predicted that by the 2030s, the effects of this occulter will have completely disappeared, allowing the full EUV flux to ionize most of the material within the Homunculus close to our LOS. Fortunately high-dispersion, STIS observations were obtained in the FUV. In the 1200 to 1700\AA\ spectra interval, most signatures of the shells, seen in absorptions of singly-ionized metals, disappeared in the FUV and nearly all of the H$_2$ at $-$513 \kms was destroyed \citep{Gull21a,Gull23}.

\subsection{Associating  velocity components with the Great and Lesser Eruptions}\label{sec:date}

The measured velocities listed in Table \ref{tab:lines} are  linear sequences  separated by significant gaps. We now estimate dates of origin by associating the two most significant absorption systems, $-$513 and $-$146 \kms, with the two historical events: the Great Eruption and the Lesser Eruption.

\cite{Smith17} measured a date of origin, 1847.1 +/- 0.8 years, for the expanding Homunculus using a 13-year baseline with \hst/WFPC2 images through the F631 filter and a 4-year baseline with \hst/ACS/HRC images through the 550M filter. This correlates well with visual photometry recorded across the Great Eruption \citep{SmithFrew11}. The $-$513 \kms\ absorption system is by far the strongest, best-defined shell. \cite{Nielsen05a} identified over 800 strong absorption lines of H$_2$ in \hst/STIS FUV spectra. \cite{Gull06} identified hundreds of absorption lines of singly-ionized metals at this radial velocity. We therefore associate the $-$513 \kms\ absorption system with the expansion date of ejection, 1847.1 +/- 0.8 years.

The importance of H$_2$ as a tracer of the Great Eruption is demonstrated by the 3D mapping of the H$_2~ \lambda2.12\mu$ emission line  (see Figure 4 in  \cite{Steffen14}).  The H$_2$ emission forms a well-defined bipolar structure. The massive ejection associated with the Great Eruption led to considerable gas that compressed to conditions ideal to the formation of H$_2$ in all directions forming the bi-lobed structure. 

Assigning an ejection date to absorption velocities of the Lesser Eruption is more challenging. \cite{Ishibashi03} discovered the Little Homunculus, a bi-lobed, ionized shell internal to the Homunculus, expanding at 170 \kms. They measured the positions of forbidden line emission knots and arrived at two estimates of origin: 1897 +/- 10 and 1904 +/- 10 years. In our LOS, this structure is associated with $-$146 \kms\ absorption.

The ejection of the slow-moving, emission-line Weigelt-clumps -- B, C, and D -- \citep{Weigelt86, Hofmann88} is associated with the Lesser Eruption. Several independent measures of the Weigelt clumps are summarized by \cite{Weigelt12}. They found the more reliable measure to be 1880 +/- 20 years. \cite{Artigau11} examined the many emission line clumps that trace an outline of a butterfly surrounding \ec.  Dates of origin for these clumps, based upon position and velocity, clustered  around the dates of the Great Eruption and the Lesser Eruption. Proper motion measures simply do not sufficiently tie down the Lesser Eruption.

The visual photometry, recorded in the nineteenth century as summarized by \cite{SmithFrew11}, indicates that the Lesser Eruption started with a rise in flux starting around 1887.2 with a plateau that extended to 1896, then a decline that extended well into the twentieth century. Assuming that the current binary orbit has not changed since that the 1890s, \cite{SmithFrew11} projected that the current binary period, 5.54 years, phased with the most current periastron, back to 1887.2, coincides with the photometric trigger point.

Implicit in this discussion is that the wind of the companion star modulates the wind(s) of the primary (merging) star across the nineteenth century. An individual shell, emerging in our LOS, formed across each periastron event leading to an absorption velocity. The measured velocities  -- rungs of the ladder -- are listed in Table \ref{tab:lines}. 

The multiple absorption velocities record changes in the wind velocity that decreases over the long term. The train of decreasing velocities is evidence that the terminal velocity is dropping with time. We likely would not see a train of increasing velocity absorptions as subsequent shells with increasing terminal velocities would overtake previous shells, given a sufficiently-long orbit. We posit that the multiple velocity absorptions are caused by passages of the current companion star through a complex, evolving envelope at various stages of the merger.

\begin{figure*}[ht]
\includegraphics[width=18. cm]{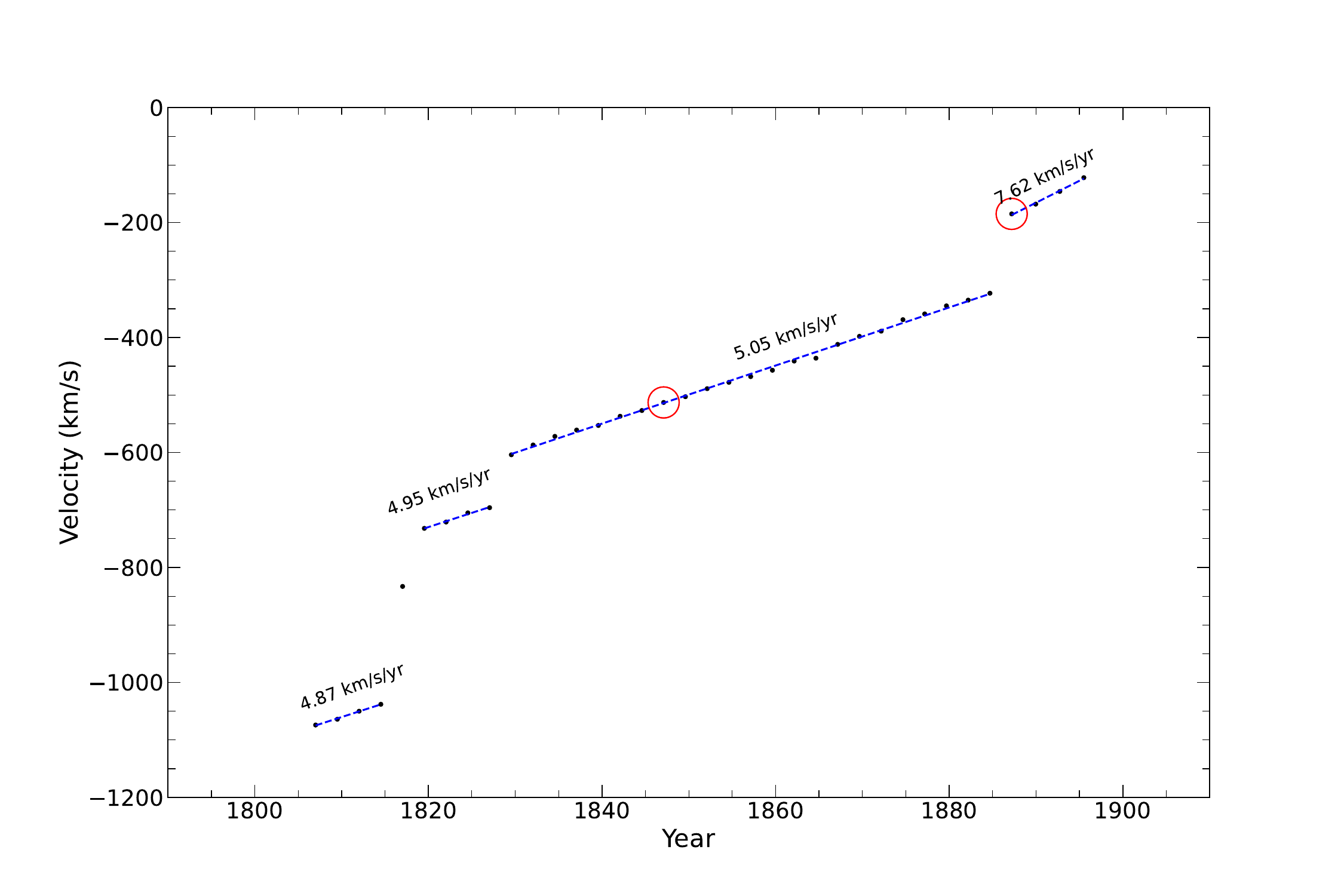}
          \caption{Observed absorption velocities plotted against estimated year of ejection based upon best information on historical observations of the Great and Lesser Eruptions. The reference dates for the x-axis are 1847.1 +/- 0.8 based upon the expansion rate of  the Homunculus using \hst\ imagery \citep{Smith17}. In our LOS, the very strong H$_2$ and metal absorptions  at $-$513 \kms\ corespond to this structure. The photometric rise of the Lesser Eruption was pegged at 1887.2 with a plateau extending through 1896 \citep{SmithFrew11}. With the assumption of a linear period between the 1847.1 and 1887.2 events, the absorption velocities are plotted against implied date. Note the apparent similarities of slope across the nineteenth century with major discontinuities. What is this telling us about the possible merger?
\label{fig:Vel}}
\end{figure*}

The four lowest absorption velocities listed in Table \ref{tab:lines},  $-$122, $-$146, $-$168 and $-$185 \kms, are separated by a large gap in velocity from the previous train of absorption velocities that end abruptly at $-$323 \kms. The LOS $-$146 \kms\ absorption velocity corresponds to the $-$170 \kms\ expansion velocity of the 3D Little Homunculus. These four velocities sample changes in the combined wind velocity across the Lesser Eruption. The photometric rise and plateau began in 1887.2 and extended to 1896 \citep{SmithFrew11}. We assign the year 1887.2 to the date of origin of the earliest absorption velocity, $-$185 \kms. Velocity changes across the plateau are sampled by $-$168, $-$146 and $-$122~\kms\ absorptions. These four shell velocities  track changes across the Lesser Eruption beginning after the major discontuity from $-$323~\kms\ to $-$186 \kms\ at the beginning of the event. A final discontinuity drops from $-$126 to $-$40 \kms, the velocity of very slow-moving ejecta, notably the Weigelt clumps \citep{Gull16}. The overall system velocity, $-$8.1 \kms, was measured of the nebular emissions of the H$_2$ and [\ion{Fe}{2}] in the infrared \citep{Smith04c}.

The orbital parameters of the binary \citep{Teodoro19} projects back to a periastron that occurred across 1887.2 $+/-$~0.1 years which correlates well with the date of the photometric rise, 1887.2 \citep{SmithFrew11}.
\subsection{An attempt to establish a binary period between the two eruptions}\label{sec:period}
Fifteen absorption velocities were identified between the $-$513 and $-$146 \kms\ absorption velocities.  An attempt to estimate a period would suggest a shell absorption event that may have occurred on average every 2.51 $+/-$~0.10 years, which is less than half of the current binary period. 
 
A first thought was to assume that the absorptions are periodic in time. Hence we apply an average interval, 2.51 years, to estimate the dates when each shell came into our LOS.

The absorption velocities in our LOS (Table \ref{tab:lines})  are plotted against implied year of absorption  in Figure \ref{fig:Vel}. The velocities bunch into four linear segments separated by large discontinuities. The first three groups, $-$1074 to $-$1038 \kms, $-$732 to $-$696 \kms, and $-$604 to $-$323 \kms, have very similar slopes. The similar slopes do not tie down the 2.51-year interval but suggest that decrease in velocity appears to be similar for for lengthy periods. Some absorptions may have been missed, or simply are not present. More likely there was an event in the system that caused a sudden drop in wind velocity. The remaining group, $-$182 to $-$122 \kms, has a noticeably shallower slope which may indicate an even slower decrease in velocity during or after the Lesser Eruption. 

The behavior of the changing velocities displayed in Figure \ref{fig:Vel} is very puzzling and appears  non-physical. Why would the velocity increments change in a linear manner from well before the Great Eruption to just before the Lesser Eruption, then shift abruptly just as there is a photometric change at the beginning of the Lesser Eruption? 

The assumption that each absorption velocity originates only from the periastron passage does not make sense. The bulk of the ejecta originated in the Great Eruption, not the Lesser Eruption. Any change in orbital period would likely occur during the merger, thought to occur during the Great Eruption. The jump from $-$323 to $-$186 \kms\ is not due to a change in orbital period.

Why are there linear segments of decreasing velocities  with large jumps between them? An alternate explanation with some physical understanding is needed.

We conclude this discussion that an orbital period close to 5.5 years is much more likely than 2.5 years.

\subsection{Absorptions created by multiple clumps or shells in a few ejections}\label{sec:few}

The multiple absorption velocities might have been caused by multiple clumps of varying opacity and ejection velocity originating from a single event. As presented in \cite{Gull23}, the spatially extended absorptions recorded as \hst/STIS CCD spectro-images with resolving power, R $=$ 10,000 and angular resolution, 0\farcs1, extended spatially across the position of \ec, well across a few arcseconds of the Homunculus. We interpret these absorptions to be from extended shells, not individual clumps. 

Additionally, it is difficult to understand the nearly linear spacing in the velocity shifts. A more acceptable interpretation is based upon what we see today, namely that the current hot secondary and its massive wind dives deeply into the extended primary wind across periastron. Analogously, in a triple star system merging into a double stellar system, the more distant star encounters portions of the distorted, merging envelope each binary period.

We suggest a more physical solution is necessary.

\subsection{Aligning the linear segments} \label{sec:linear}
\begin{figure*}
    \includegraphics[width=18. cm]{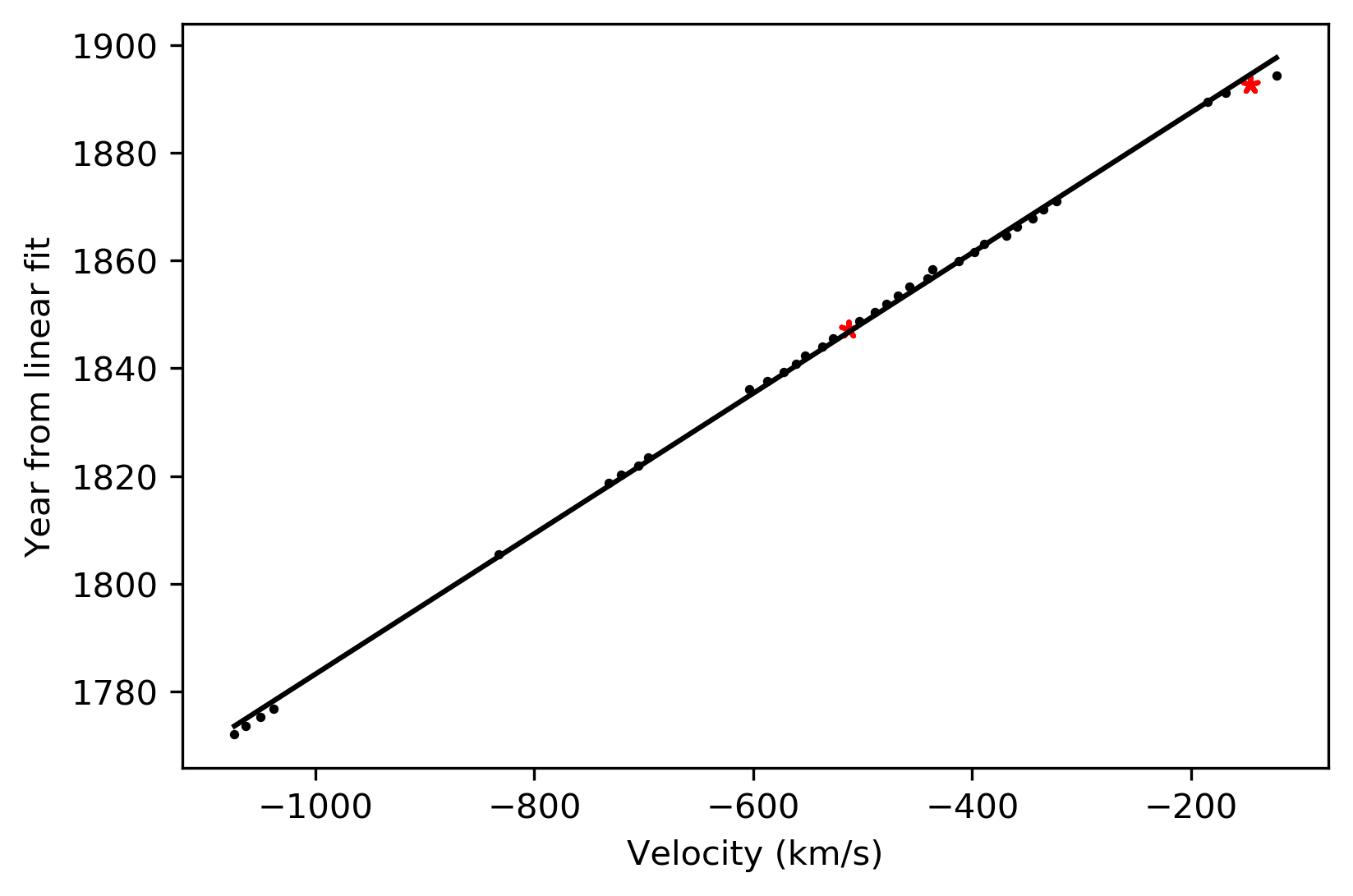}
          \caption{Absorption velocities now plotted by shifting the segments to form a continuous line. Each of the four groups of velocities are shifted independently. The dates of origin now extend to the 1770s which are similar to dates of origin measured of \ion{Mg}{2} emission clumps external to the Homunculus based upon two epochs of \hst/ACS images \citep{Morse24}.
\label{fig:linear}}
\end{figure*}

Our LOS passes through the wall of the Homunculus well away from its axis of symmetry. \cite{Madura10, Madura12} modeled the colliding wind structures to interpret high-spatial-resolution mappings recorded by \hst/STIS of forbidden emissions projected within an arcsecond of \ec.  They determined that the axis of the binary orbital plane was closely aligned to the Homunculus axis of symmetry and that the periastron passage of \ec-B was on the far side of \ec-A.

We consider an alternate approach to understand the velocity ladder. Perhaps the large changes in velocities between the linear stretches are due to no shells in our LOS for a period of time. The Homunculus has a large, continuous, well-defined bipolar structure as  mapped by the H$_2$ 3D structure \citep{Steffen14}. We know that the bulk of the dust appears to be in the orbital plane on the far side of \ec\ \citep{Morris17, Morris20}, where also a large structure seen in NH is present as mapped by the {\it Herschel Observatory} \citep{Gull20}. The equivalent widths of absorption profiles decrease from that of $-$513 to $-$323 \kms\ then disappear until an absorption appears at $-$186 \kms\ as shown in Figure \ref{fig:spect} and multiple examples shown in \citep{Gull06}.

The velocity increment appears to be continuous from well before the Great Eruption to well after the Great Eruption. Hence we assume that the wind velocities generated by the system across the Great Eruption was a continuous mass loss process. Despite the huge increase in mass loss at one specific time, the wind velocities changed continuously. 

We use the segmented, linear behavior of the terminal winds and shift the segments to align with a constant slope as displayed in Figure \ref{fig:linear}. The aligned segments suggest that shells formed in our LOS as early as the 1760s, that there was a wind velocity that systematically decreased across nearly 130 years, but that there were intervals where shells did not extend across our LOS. This may agree with the highly variable equivalent widths of the absorption profiles seen across the high- and low-ionization states. Such might be the result of the mass in the winds decreasing or the orientation of the colliding winds changing with time.

The ladder velocities are from only one narrow sample through the ejecta (Figure \ref{fig:linear}) and show  similar dates of origin of shells, 1760 to 1847, of shells with velocities more negative than $-$ 513 \kms. The less negative velocities originate from shells within the Homunculus, thus extending the record of the wind structures associated with the Great Eruption to dates closer to the Lesser Eruption.

The several gaps in absorption velocities provide clues to the shapes of the individual shells. Likely the shells are densest in the orbital plane, but may drop off well before the polar region. Two effects may contribute to the lack of shells at large angles out of the plane: less material flows towards the orbital pole and more EUV radiation escapes to the orbital pole. The latter would lead to far less momentum transfer as very few resonant transitions occur in the EUV compared to the NUV. Hence  wind flows toward the polar regions would not evolve narrow absorption profiles.

If the shells formed from a binary wind interaction, as shells appear to form in recent times \citep{Teodoro16}, then we would expect to see eight absorption velocities evenly distributed between the $-$513 and the $-$185 \kms\ absorptions. Fifteen were identified plus a large gap of 138 \kms\ just before the $-$185 \kms\ absorption identified with the beginning of the Lesser Eruption. 

Likewise, two gaps in velocity occur between $-$ 684 and $-$ 604 \kms\ and between $-$1038 and $-$832 \kms\ (ignoring the $-$833 \kms\ absorption). Apparently something led to shells not being in our LOS before the advent of the Great Eruption.

We also note that the equivalent widths of the absorption velocity components decrease from $-$513 to $-$323 \kms\ which would be expected if the column densities of the shells decrease with time.

\begin{figure}
\includegraphics[width=8.5cm]{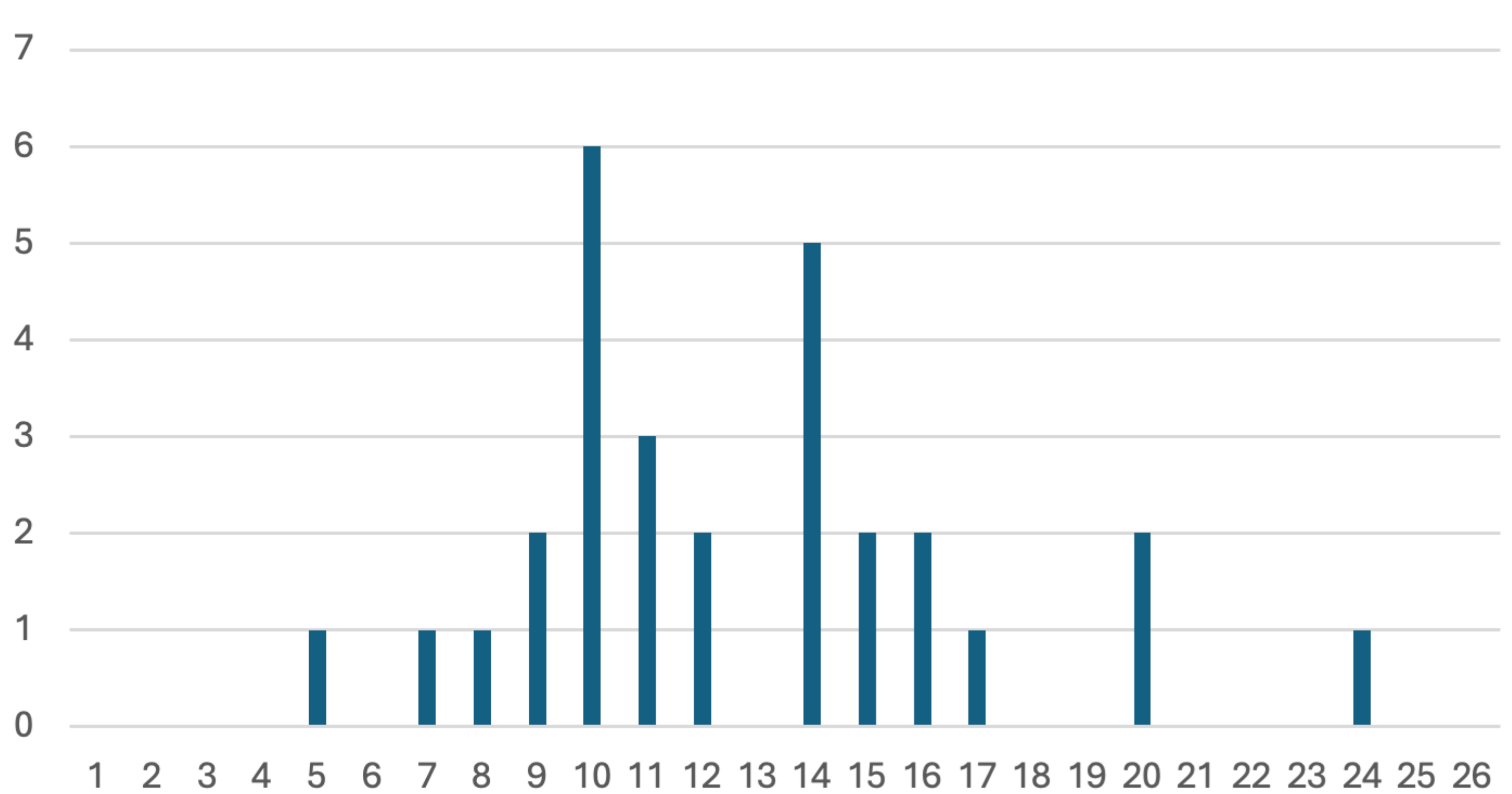}
          \caption{Frequency of velocities extending from before the Great Eruption to afterwards. The two most populated velocities, 10 and 14 \kms\ may be modulated substructure of the merger. Estimated errors of measure are one to two \kms.}
\label{fig:freq}
\end{figure}

In an attempt to look for periodicity, we examined the frequency of velocity shifts from $-$1074 to $-$323 \kms\ as displayed in Figure \ref{fig:freq}. The two most frequent changes in velocity were 10 and 14 \kms, but there is a large dispersion of very low and very high velocity changes, ignoring the large discontinuities mentioned above. 

Recently \cite{Morse24} presented expansion measures of outer ejecta using \hst/WFC3 images of \ion{Mg}{2} $\lambda$2800 recorded in 2018 and 2020. They found about half of the clumps moved at transverse velocities consistent with the 1847.1 origin of the bipolar Homunculus as previously measured by \cite{Smith17}. The structures measured by \cite{Morse24} are part of or external to the Homunculus.  The transverse velocities plotted in their Figure 2 demonstrate a certain symmetry of the SE and NW lobes. The  other half, located outward of the Homunculus, had positions and transverse velocities consistent of origins distributed before 1847 as early as 1760. Such suggests that the merger extended over many decades, not as a singular event.

The dates of origin, inferred from the linear change in velocity with time as plotted in Figure \ref{fig:linear}, are listed in Table \ref{tab:lines}, right hand column. Notably the dates of origin extend from 1847.1 to the 1770's, which trends similar to the dates of origin of the emission structures measure by \cite{Morse24}.

\subsection{Could the shells be due to DACs?}\label{sec:dac}
Discrete absorption components, or DACs, are seemingly ubiquitous in massive hot stars \citep{Howarth89}. These are the observed effects of co-rotating interaction regions, or CIRs, in the winds as modeled by \cite{Cranmer96}. These are often seen in ultraviolet resonance lines for main-sequence O stars, as well as sometimes seen in evolved massive stars. The DACs are seen to start at low velocities and propagate to the terminal wind speed  fairly broad at first, becoming ever narrower as V$_{inf}$ is approached. Time-series of ultraviolet wind lines of luminous blue variables are not as frequently seen as those of other hot stars, owing in large part to their higher amount of interstellar extinction than many OB stars thus making them harder to observe with IUE in the past. However, \cite{Richardson11} presented an intense time-series of optical H$\alpha$ spectroscopy of the LBV P Cygni, the only other Galactic LBV with an observed eruption. With more than a decade of uniform instrumentation, they found a DAC-like feature with a long recurrence time of 1700 days, much longer than the expected rotation period of the star.

With eta Car, we can expect that the velocity shells we observe were from material ejected during the Great Eruption. One possible way to explain these shells would be to assume that these were similar DACs as those seen in the H-alpha line of P Cyg reported by Richardson et al. (2011). In this case, we could envision that the terminal velocity of the very hot, recently merged object would be at -1000 km/s or similar, but the newly merged object could start to expand and cool. Thus, each DAC would have a different terminal velocity, resulting in the half dozen or so shells we see at very high velocity. Then, as the merger progresses, the star will be puffed up to show a lower effective temperature, which in turn provides a lower terminal velocity. Now our DACs will have much lower terminal speeds and hence velocities in today’s observed shells than those in the early parts of the eruption. This much lower effective temperature is supported with the spectral properties of the Great Eruption’s light echoes \citep{Prieto14,Rest12}. As the star continues to cool and reach an equilibrium, the DACs build up shells of various column density based on the mass-loss rate that was likely not constant during the eruption. A similar scenario could have existed during the Lesser Eruption to account for the relatively low velocities in the family of absorptions. 

 While the concept of DACs or CIRs cannot be completely dismissed, the scale of the material in the shells appears to be much larger than the typical DAC. We remind the reader that the current binary system produces shell-like structures during each periastron passage \citep{Teodoro16} which is a very different mechanism that those which produce DACs and CIRs. 

\subsection{Associating the velocity ladder with the merger scenario. \label{sec:scenario}}
  
\begin{figure}[ht]
\includegraphics[width=8.5 cm]{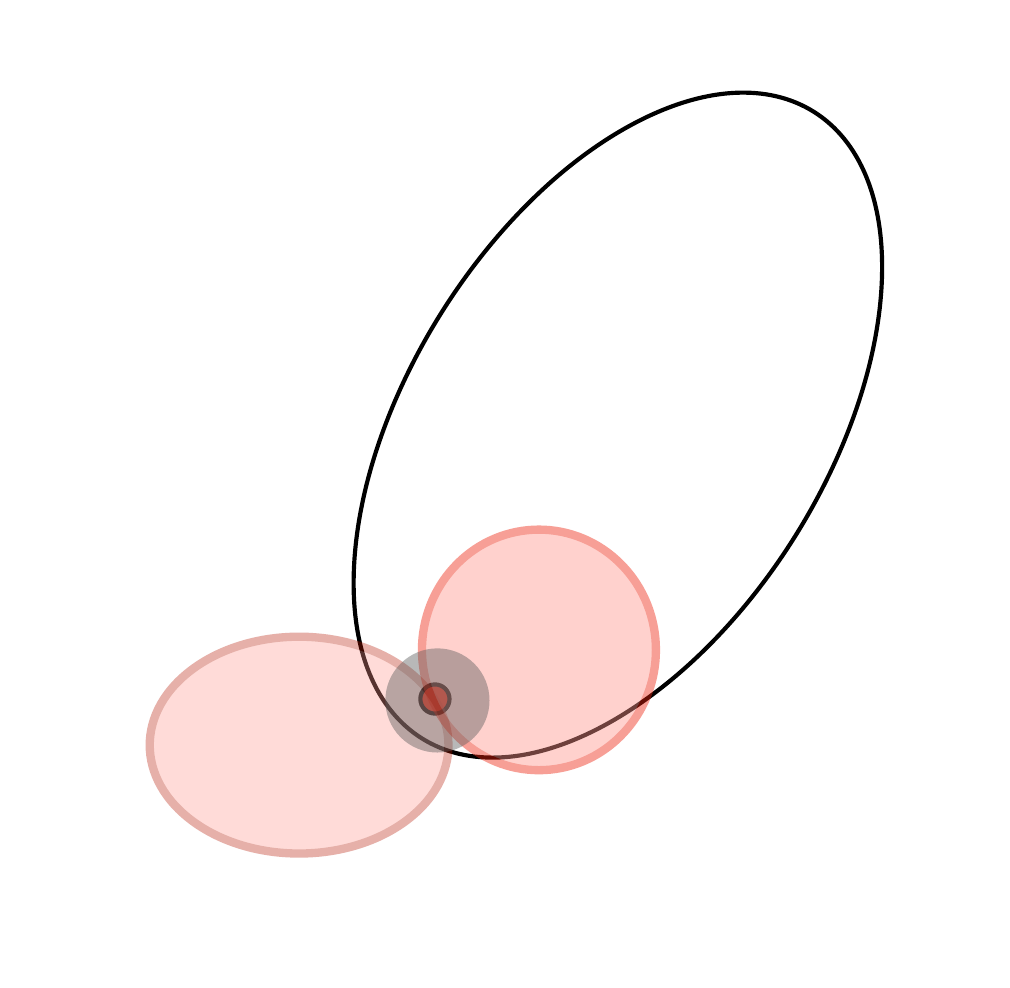}
          \caption{Transitional, bi-lobed common envelope structure from the Great Eruption to the Lesser Eruption. The cores of two massive stars have merged during the Great Eruption, while the extended atmospheres formed a distorted, bi-lobed envelope that gradually conforms to the newly formed core. The Lesser Eruption appears to be a final disgorgement of the common envelope leading to the current, very extended primary wind of the current binary. (Concept adapted from \cite{Hirai21}, Figure 1) 
\label{fig:trans}}
\end{figure}

 We now consider how this absorption ladder may support a binary merger in a triple system. Alternative scenarios may explain the ladder of absorptions. Indeed we encourage others to suggest other viable models that explain the observations!
 
Concepts of a triple system explaining the behavior of \ec\  go back as far as \cite{Livio98}  who suggested that the current \ec\ is a triple system. However \cite{Teodoro16} found that the present-day behavior of \ec\ is well explained by the binarity discovered by \cite{Damineli96}. \cite{Portegies16} provided a computationally intensive model of a binary merger within a triple system that addressed the Great Eruption and  the current binary system. However more detail was necessary to compare theory to observtations.

Recently \cite{Hirai21}  investigated triple systems that evolved by mergers to binaries with ejecta as in the \ec\ and Homunculus system. Their Figure 1 summarizes the various stages leading to the merger: a relatively stable triple system that lasts for several million years where the member stars basically evolve independently; an unstable triple with a thousand to ten thousand year interval as one or more stars swell up and begin to interact, leading to a chaotic series of interactions; the merger period where two evolved stars spiral into each other initially creating a common envelope with much material flowing outward leading to the actual merger of two stars. Finally the wind-ejecta interaction stage begins with the merged star and the remaining companion interacting with each other and influencing the ejecta.

We focus on the last stage described by the \cite{Hirai21} model: the core merger of two stars that led to settling surrounded by a bi-lobed envelope while the third star orbits around the merging structure. We present a sketch of this stage in Figure \ref{fig:trans}. \cite{Hirai21} suggested that the settling time would be decades to a century. This timescale is comparable to the period covered by the velocity ladder which begins decades before the merger (the Great Eruption) and lasts through the final settling  (the Lesser Eruption) of the distorted, bi-lobed, common envelope of the merged stars. As the two stars approach the core-merger event, a common envelope develops that persists well beyond the actual core merger. 

\cite{Hirai21} suggested that the common envelope gradually dissipates but persists for up to a hundred years. The rungs of velocity ladder, plotted in Figure \ref{fig:linear}, would be generated by the third star passing through the complex, shared envelope of the merging binary with a period close to 5.5 years (Figure \ref{fig:trans}). Throughout most of the period of the third star orbiting around the merging system, its wind clears out the shared envelope. Each time that the third star encounters the merging, rotating structure, a shell of gas moves toward our LOS and is compressed by colliding winds and radiation.  

However the merging structure was a highly distorted structure with extended ears beyond the newly-core star. The third star, now the binary companion, would always pass through the greatly extended envelope of the merged star. Additionally the companion star might pass through the extended lobes resulting from the merger as the extended envelope slowly settles to a new normal. If the merging stars had been co-planar with the third star, then the companion might pass through both lobes left over from the merger. 

Possible evidence of the three stars being nearly co-planar in orbit may be the nearly symmetrical bipolar shape of the H$_2$ structure traced out by \cite{Steffen14} as displayed in their Figure 4. Of note are two 'ears' that protrude from the bipolar structure: one extends to the NE and the other to the SW. Such may be due to small assymetries of the merger, but notable is the fact that the axis of symmetry of the Homunculus is closely aligned to the orbital pole of the current binary. Such strongly suggests that the merging binary had a orbit that at the time of merger was close to the orbital plane of the binary plus third star.

The velocity ladder appears to be a record across the nineteenth century of the merged star and common envelope being modulated by the fast wind of the current secondary star. The wind of this secondary star interacted with the merged envelope and merged star, interrupting the slow-moving gas across the bulk of the orbit. In analogy to the modern-day binary system \citep{Madura13}, the third star during its orbit first dove through the bipolar envelope then the core atmosphere and through the opposite portion of the bipolar envelope potentially interrupting the wind flows three times forming up to three shells each 5.5-year orbit (see Figure 10 in \cite{Madura13}. 

The merger apparently took place across several cycles in the 1840s with the bulk of the ejecta occurring in 1847. The common envelope existed well before that time, possibly as early as the 1760s. The gaps in velocity may be a record of times when abrupt changes occurred as the two merging stars spiraled inward. The gaps, between $-$1038 and $-$732 \kms\ and $-$696 and $-$604 \kms, may be caused by the chaotic behavior of the triple system prior to the merger suggested by \cite{Hirai21}. It does seem peculiar that the gaps are seen as drops, not increases in expansion velocities. However shells with increasing velocities with time would overtake earlier shells and wipe them out.

The slope, defined by the four lowest velocities, is less than the slope defined by the higher velocity segments. One interpretation is that the slope, defined by these four lowest velocities, may be the final settling of the common envelope during the Lesser Eruption.

The jump from $-$323 to $-$185 \kms\ appears to be associated with the event corresponding to the Lesser Eruption. Only four shell velocities were recorded below $-$323 \kms\  and may indicate the final dissipation of the merged envelope by 1895. The origin of the Weigelt clumps is consistent with the 1890s; they likely are the final ejection of the merged envelope lifting off at very slow velocities estimated to be about $-$40 \kms.

Our LOS views the current orbital plane tilted at 45$^o$. Given the chaotic nature of a ninteenth-century triple system, intermittent gaps in the shells that we view would be expected.

In the meantime, the bipolar envelope from the merged binary would be gradually relaxing, but in a chaotic way. The observed velocity ladder suggests a relaxation time of 40 to 50 years from the actual merger until the last major wind flow comes out in the form of very slowly-moving clumps.

One puzzlement is the apparent linear decrease in wind velocity with time. The Great Eruption led to about 40 \Msun\ being ejected. A change in velocity increment might be expected at the time of merger. However, we note that mass loss, while  most at the time of the Great Eruption, appears to span well before and well after the merger. 

A second puzzlement is that we apparently do not detect each shell caused by passage of the secondary through the extended primary wind. We should expect to see nine absorption velocities at equal intervals between $-$513 and $-$146 \kms. We do not. Could this be evidence that the primary envelope shrunk to below 2 au diameter for several cycles preceding the Lesser Eruption? The final drop in velocity from $-$126 to $-$40 \kms\ may support a temporarily shrinking of the newly-formed primary structure. Perhaps accelerated shells did form briefly late in the velocity gap and wiped out some slower, previously-formed shells. 

Is there an alternate explanation?

\section{Conclusions}\label{sec:con} 

A series of 38 narrow-absorption velocities, measured in our LOS from \ec\ provides a record of wind velocities that decreased with time. Referencing two key absorption velocities to the Great and Lesser Eruptions, we find this velocity ladder spans from the 1890s possibly back to the 1770s. 

However, this velocity ladder is only a partial record of events that occurred during the 18th and 19th centuries. While we find many possible interpretations, the simplest interpretation appears to be that the absorptions rise from shells that record interactions of a companion star to a merging binary. Analogous to Occam's razor, this may sever the Gordian knot of complexity that is the ejecta formed from major events.

The velocities, more negative than $-$513 \kms\ which correlates with the 1847.2 measure of the Great Eruption \citep{Smith17}, suggest dates of origin that are similar to dates of origin of emission structures determined by \cite{Morse24} to be as early as the 1760s.

The velocity ladder includes velocities less negative than $-$513 \kms\ which leads to shells originating later than 1847.2, into the 1860s but well before the start of the Lesser Eruption in 1887.2

Large discontinuities exist between  sequences of velocities that show a near linear decrease in velocity with time. We suggest these discontinuities in our LOS are due to weakening of the wind velocities and mass loss of \ec\ with time. Offsetting the sequences of velocities to form a straight line in velocity increments leads to inferred dates of origin that extend back to the 1760s and forward to the 1890s. 

The frequency of the absorption velocities exceeds that which would be expected from the secondary star passing through the extended primary envelope each orbit. Instead, placing all of the velocities on a linear curve, suggests that the secondary passed through the preceding and following ears of the structure caused by the merger. 

This velocity ladder, considered in context of the well-defined bipolar H$_2$ shell \citep{Steffen14} with minimal asymmetries, suggests that the three stars across this interval of time were nearly co-planar in orbit.

The current binary is stable on the scale of decades as witness the half-century calorimetry measures and the repeatability of the X-ray light curve. The limited FUV observations with the \hst/STIS are also consistent with a stable binary system with no evidence for changing stellar parameters over recent decades.

Two lingering puzzles are 1) the linear decrease with time of the shell velocity and 2) the absence of an absorption velocity caused by periastron passages in  several orbits preceding the Lesser Eruption. 

Further studies and models are necessary to test this suggestion that the ladder of velocities are a record of a merging binary in a triple system. 

Additional monitoring with the UV \hst/STIS, especially during both the high- and the low-ionization periastron passages, is needed to follow the ladder velocities to determine how excitation, ionization, and  the measured velocities evolve with time. As the apparent UV fluxes of \ec\ have increased ten-fold in our direction over the past two decades, even higher resolving power spectra with high S/N are possible and must be pursued. Many singly-ionized metal absorptions will disappear during the high-ionization state only to reappear during the low-ionization state. Additional absorption velocities may appear in our LOS.

Observations in the high-ionization state are likewise needed to follow the evolution of the shells as the extreme UV continues to increase. However the singly-ionized metal resonant absorptions, accessible in the MUV and NUV, are being replaced by doubly-ionized metal absorptions located shortward of Ly~$\alpha$, a spectral interval that was found to be  very complex as shown by the {\it FUSE} observatory with its limited spectral and angular resolutions \citep{Iping05a}.

New insight will also come from {\it JWST} and {\it ALMA} observations capable of penetrating the dust  and parsing out the molecular and neutral gas structures.

\begin{acknowledgments}This discussion was not funded by any facility. Many thanks by the primary author to the Eta Carinae ZOOM team and the numerous STScI personnel that assisted in the original observations. The material is based upon work supported by NASA under award number 80GSFC21M0002.
\end{acknowledgments}

\facility{HST(STIS)}

\bibliography{ref}{}
\bibliographystyle{aasjournal}


\end{document}